\documentclass[conference]{IEEEtran}
\IEEEoverridecommandlockouts
\usepackage{mathrsfs}
\usepackage{amsfonts}
\usepackage{ifpdf}
\usepackage{cite}
\usepackage{array}
\usepackage{booktabs}
\usepackage{setspace}

\ifCLASSINFOpdf
\usepackage[pdftex]{graphicx}
\else \fi
\usepackage[cmex10]{amsmath}
\usepackage{array}
\usepackage{mdwmath}
\usepackage{mdwtab}
\usepackage{eqparbox}
\usepackage[tight,footnotesize]{subfigure}
\usepackage{algorithmic}
\usepackage{algorithm}
\usepackage{amsmath}
\usepackage{epsfig}
\usepackage{ae}
\usepackage{amssymb}
\usepackage{times}
\usepackage{amsmath}   

\usepackage{xcolor}

\usepackage{amsthm} 

\usepackage{supertabular} 

\usepackage{multirow}
\newlength\savewidth
\newcommand\shline{\noalign{\global\savewidth\arrayrulewidth
                            \global\arrayrulewidth 1.5pt}%
                   \hline
                   \noalign{\global\arrayrulewidth\savewidth}}

\begin{document}

\title{
Inter-Carrier Interference Mitigation for\\ Differentially Coherent Detection in\\ Underwater Acoustic OFDM Systems
\thanks{This work was supported in part by the National Science Foundation of China under Grant 61971462, Grant 61831013, and Grant 61631015 and the State Key Laboratory of Integrated Services Networks (Xidian University) under Grant ISN19-09.}
}
\author{
    \IEEEauthorblockN{
        Yunlong Huang and Yuzhou Li
    }
    \IEEEauthorblockA{
        Huazhong University of Science and Technology, Wuhan, 430074, P. R. China\\
        State Key Laboratory of Integrated Services Networks (Xidian University), Xi'an, 710071, P. R. China\\
        ylhuang@hust.edu.cn, yuzhouli@hust.edu.cn}
    }
\maketitle
\IEEEpeerreviewmaketitle
\begin{abstract}
Suppressing the inter-carrier interference (ICI) is crucial for differentially coherent detection in underwater acoustic (UWA) orthogonal frequency division multiplexing (OFDM) systems due to the fact that the UWA channel is inherently violently Doppler-shifted. In this paper, we propose a new ICI suppression method, referred to as the partially-shifted fast Fourier transform (PS-FFT), which eliminates the ICI from both the time and frequency domains. Specifically, the PS-FFT first divides the received signal in the entire block duration into several short non-overlapping ones to reduce the channel variation in the time domain. It then applies the Fourier transform at several predefined frequencies to the received signal in each of these intervals to compensate Doppler shifts in the frequency domain. Finally, it weightedly combines the multiple demodulator outputs at each carrier as one output for symbol detection, with the combiner weights being solved by the stochastic gradient algorithm. Simulation results show that the PS-FFT dramatically outperforms the existing classical methods, the partial fast Fourier transform (P-FFT) and the fractional fast Fourier transform (F-FFT), for both medium and high Doppler factors and large carrier numbers in terms of the mean squared error (MSE). Numerically, the MSE of the PS-FFT is reduced by $\bf{61.83\%-84.89\%}$ compared to that of the F-FFT when the input signal-to-noise ratio (SNR) at the receiver ranges from 10 dB to 30 dB at a Doppler factor of $\bf{3\times 10^{-4}}$ and a carrier number of 1024 where the P-FFT even cannot work.
\end{abstract}
\section{Introduction}
Orthogonal frequency division multiplexing (OFDM), attributed to its powerful abilities in combating the frequency selectivity and improving the bandwidth efficiency, has been recognized as a prospective technique for high-data-rate underwater acoustic (UWA) communications \cite{EditorialUWACommunications}. In order to recover data symbols from the received OFDM signal, conventional systems usually adopt the coherent detection technique at the receiver \cite{MarineWirelessBigData,ToRelayOrNotRelay,ASurveyofUnderwaterMagnetic}. Nevertheless, one of the critical problems for accurate data recovery with respect to the coherent detection is the requirement of the precise channel estimation. To this end, it is needed to insert a certain number of pilots into each OFDM block to assist the receiver to obtain the channel state information (CSI). This, however, would inevitably consume the extremely limited frequency resources of the UWA channel, typically only on the order of kHz to tens of kHz \cite{ProspectsProblemsofUWAComm}, and thus substantially reduces the bandwidth efficiency and the data rate.

Contrasted to the coherent detection, the differentially coherent detection has the potential to eliminate the requirement for channel estimation, and hence draws significant attentions recently \cite{MS_MFFF1_PFFT_Diff_Nomultichan,MS_MFFT8_PFFT_Diff_Multichan,JH_MFFT_PFFT_Eigen,MS_MFFT3_FFFT_Diff_Multichan,MS_MFFT5_MFFT_Diff_Multichan}. Considering an OFDM system with differential encoding in the frequency domain, the channel frequency response can be regarded to vary slowly between adjacent carriers as long as the carrier spacing $\Delta f$ is small enough. Utilizing this fact, the data symbols can be detected differentially without the CSI at the receiver. Moreover, the detection performance can be further improved by increasing the number of carriers $K$, as the coherence between adjacent carriers is enhanced when the carrier spacing $\Delta f=B/K$ becomes narrower, where $B$ denotes the bandwidth. Most importantly, increasing the number of carriers would also improve the bandwidth efficiency of the OFDM system, given by $R/B\sim \mathrm{log_2}M /(1+T_{g}B/K)$, where $R$, $M$, and $T_g$ are the bit rate, the modulation order, and the guard interval between blocks, respectively. However, shortening the carrier spacing $\Delta f$ would in turn make the OFDM system quite sensitive to the frequency offset, and thus the detection performance possibly significantly drops without substantial inter-carrier interference (ICI) mitigation as the UWA channel is inherently violently Doppler-shifted \cite{EditorialUWACommunications,MarineWirelessBigData,ASurveyofUnderwaterMagnetic}. Therefore, it is important to use as many carriers as possible to insure strong coherence between adjacent carriers and also high bandwidth efficiency and meanwhile maintain the ICI at a low level in differential OFDM systems.

A series of approaches that concentrate on the ICI suppression before the fast Fourier transform (FFT) demodulation have been proposed in OFDM systems \cite{MS_MFFT2_PFFT_Nodiff_Nomultichan,MS_MFFF1_PFFT_Diff_Nomultichan,MS_MFFT8_PFFT_Diff_Multichan,JH_MFFT_PFFT_Eigen,MS_MFFT3_FFFT_Diff_Multichan,MS_MFFT5_MFFT_Diff_Multichan,MX_MFFT_DFFFT}. In general, these methods can be grouped into two categories, with the partial fast Fourier transform (P-FFT) and its revised versions being the first category that mitigate the ICI from the time domain \cite{MS_MFFT2_PFFT_Nodiff_Nomultichan,MS_MFFF1_PFFT_Diff_Nomultichan,MS_MFFT8_PFFT_Diff_Multichan,MS_MFFT5_MFFT_Diff_Multichan,JH_MFFT_PFFT_Eigen} and the fractional fast Fourier transform (F-FFT) and its revised versions being the second one from the frequency domain \cite{MS_MFFT3_FFFT_Diff_Multichan,MS_MFFT5_MFFT_Diff_Multichan,MX_MFFT_DFFFT}. Specifically, the P-FFT, first proposed in \cite{MS_MFFT2_PFFT_Nodiff_Nomultichan}, divides the block duration into several non-overlapping intervals to reduce the channel variation. Performing the Fourier transform to each of these intervals yields multiple outputs at each carrier, which are then weightedly combined as one output for symbol detection. In \cite{MS_MFFF1_PFFT_Diff_Nomultichan}, the P-FFT was first investigated in differentially coherent detection systems, and a stochastic gradient algorithm (SGA) was developed to solve the combiner weights. Further, \cite{MS_MFFT8_PFFT_Diff_Multichan} extended the differential system with only one receiving element in \cite{MS_MFFF1_PFFT_Diff_Nomultichan} to the system equipped with multiple receiving elements. An eigendecomposition-based algorithm for weight determination was devised in \cite{JH_MFFT_PFFT_Eigen} to avoid the sensitivity of the SGA to the initial points. The F-FFT, first proposed in \cite{MS_MFFT3_FFFT_Diff_Multichan}, carries out the Fourier transform at several different frequencies, including the carrier frequency $f_k$ and frequencies offsets to $f_k$ by fractions of the carrier spacing $\Delta f$, to compensate Doppler shifts, and then combines the multiple outputs as that in the P-FFT \cite{MS_MFFF1_PFFT_Diff_Nomultichan,MS_MFFT8_PFFT_Diff_Multichan}. In \cite{MS_MFFT5_MFFT_Diff_Multichan}, a gradient scaling and a thresholding method were introduced to improve the SGA in \cite{MS_MFFF1_PFFT_Diff_Nomultichan,MS_MFFT8_PFFT_Diff_Multichan,MS_MFFT3_FFFT_Diff_Multichan}. Furthermore, \cite{MX_MFFT_DFFFT} developed the decision fractional fast Fourier transform (DF-FFT) to improve the performance of the F-FFT through symbol rebuilding.

It should be pointed out that the performance achieved by differentially coherent detection with the P-FFT \cite{MS_MFFF1_PFFT_Diff_Nomultichan,MS_MFFT8_PFFT_Diff_Multichan,MS_MFFT5_MFFT_Diff_Multichan} or the F-FFT \cite{MS_MFFT3_FFFT_Diff_Multichan,MS_MFFT5_MFFT_Diff_Multichan} significantly outperforms conventional detection methods with a single FFT. However, the mean squared error (MSE) gets worse rapidly when Doppler factors are on the order of $10^{-4}$, typical values in real UWA environments, and the number of carriers is larger than 1024. Using few carriers may improve the MSE performance, but it in turn not only reduces the coherence between adjacent carriers but also decreases the bandwidth efficiency. In other words, the improvement of the differentially coherent detection with the P-FFT or the F-FFT is still limited, especially in the case of large Doppler
factors and carrier numbers.

In view of these, this paper extends the existing time- and frequency-domain methods to the time-frequency domain to develop a new ICI mitigation method, referred to as the partially-shifted fast Fourier transform (PS-FFT), which works well even when the Doppler factor and the number of carriers are both large. Specifically, the entire block duration is first divided into several short non-overlapping intervals for channel variation reduction. Then, the Fourier transform is applied at several predefined frequencies to the received signal in each of these short intervals to compensate Doppler shifts, yielding multiple demodulator outputs at each carrier. By these, the PS-FFT simultaneously takes advantage of the superiority of the P-FFT and the F-FFT in mitigating ICI effects. Finally, it solves the combiner weights for combining the multiple outputs by utilizing the SGA as in \cite{MS_MFFT5_MFFT_Diff_Multichan}. Simulation results show that the PS-FFT significantly outperforms the P-FFT and the F-FFT for both medium and high Doppler factors and large carrier numbers in terms of the MSE.

The remainder of this paper is organized as follows. In Section~\ref{Section:SystemModel}, we introduce the system model for the UWA differential OFDM system. Section~\ref{Section:PSFFTDiffDetection} describes the detailed process of the proposed PS-FFT demodulation for differentially coherent detection. Simulation results are presented in Section~\ref{Section:SimulationResults} to verify the performance of the proposed method. Finally, conclusions are summarized in Section~\ref{Section:Conclusions}.

\section{System Model} \label{Section:SystemModel}
We consider a differential OFDM system with $K$ carriers, in which the data symbol $d_k$ transmitted on each of the carriers is obtained by applying the differential encoding to the original data symbol $b_k$ using the same encoding scheme as in \cite{MS_MFFF1_PFFT_Diff_Nomultichan,MS_MFFT8_PFFT_Diff_Multichan,JH_MFFT_PFFT_Eigen,MS_MFFT3_FFFT_Diff_Multichan,MS_MFFT5_MFFT_Diff_Multichan}, given by
\begin{equation} \label{Eq:DifferentialEncode}
d_k=
\begin{cases}
b_{k}d_{k-1}, & {1 \leq k \leq K-1}\\
a_0,  & {k=0}
\end{cases}
\end{equation}
where $b_k$ is generated from the $Q$-ary unit-amplitude phase-shift keying (PSK) constellation alphabet set $\mathcal{A}=\left \{a_0,a_1,\ldots,a_{Q-1} \right \}$, in which the constellation symbol $a_q=e^{j2\pi q/Q},q=0,1,\ldots,Q-1$.

After applying the inverse Fourier transform to these differentially encoded symbols $d_k$, the time-domain transmitted signal $s(t)$ is calculated as
\begin{equation} \label{Eq:TransmitSignal}
s(t) = {\rm Re}\left \{\sum_{k=0}^{K-1}d_{k}{e}^{j2\pi f_{k}t}\right \},\ \ t\in [0,T]
\end{equation}
where $f_k=f_0+k\Delta f$ is the carrier frequency corresponding to the $k$-th carrier and $T=1/\Delta f$ is the block duration.

Transmitting the signal $s(t)$ through a multipath UWA channel with the path gain $h_p$ and path delay $\tau _p$ corresponding to each path, the received signal $r(t)$ is represented as
\begin{equation} \label{Eq:ReceiveSignal}
r(t) = \sum_{p=0}^{P-1} {h_{p}(t)s(t-\tau _{p}(t))}+n(t)
\end{equation}
where $n(t)$ is the noise at the receiver.

Finally, after carrying out a series of process on the received signal $r(t)$, including frame synchronization, initial resampling, downshifting by the lowest carrier frequency $f_0$, and removal of the guard interval, we will obtain the corresponding baseband signal $v(t)$, which is modeled as
\begin{equation} \label{Eq:BasebandReceiveSignal}
v(t) = \sum_{k=0}^{K-1} {H_{k}(t)d_{k}e^{j2\pi k\Delta ft}}+u_{k}(t)
\end{equation}
where $H_k(t)=\sum_{p=0}^{P-1} {h_{p}(t)e^{-j2\pi f_{k}\tau _{p}(t)}}$ and $u_{k}(t)$ are the channel coefficient and the equivalent noise corresponding to the $k$-th carrier, respectively.

\begin{figure}[t]
\centering \leavevmode \epsfxsize=3.5 in  \epsfbox{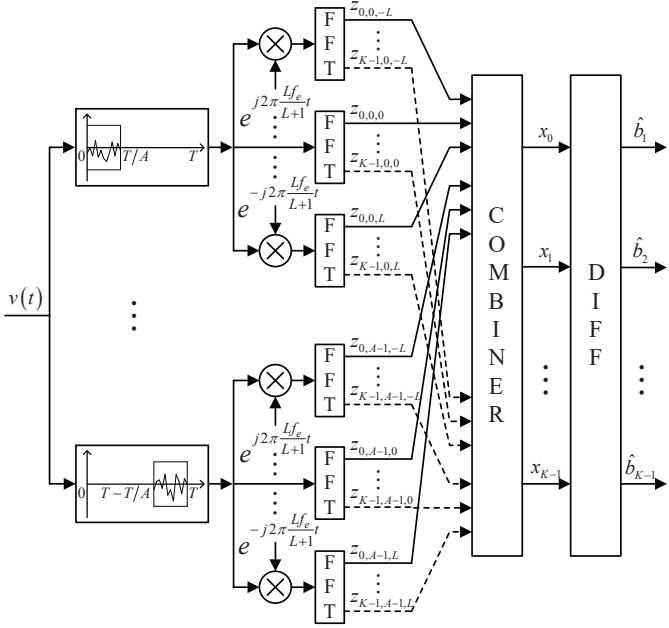}
\centering \caption{Block diagram of the PS-FFT demodulation where ``DIFF'' denotes differentially coherent detection.} \label{Fig:PSFFT_Demodulation}
\end{figure}

\section{PS-FFT Demodulation for Differentially Coherent Detection} \label{Section:PSFFTDiffDetection}
In this section, we first describe the process of the conventional differentially coherent detection scheme with a single FFT, and analyze its problems in the UWA system with serve Doppler shifts. Then, the PS-FFT and its process for differentially coherent detection are introduced in detail, including mitigating the ICI from both the time and the frequency domains, weighted combination, symbol detection, and the weight updating algorithm.
\subsection{Conventional Differentially Coherent Detection with a Single FFT Demodulation} \label{Subsection:SingleFFTDemodulation}
In the conventional differentially coherent detection scheme, the Fourier transform is applied to the signal $v(t)$ directly, yielding the received symbol $x_k$ at each carrier as
\begin{equation} \label{Eq:SingleFFTDemodulation}
\begin{aligned}
x_k = \int_{0}^{T}{v(t)e^{-j2\pi k\Delta ft}dt},\ \ k=0,\ldots,{K-1}.
\end{aligned}
\end{equation}

When the channel variation is negligible, the relationship between the received symbol $x_k$ and the transmitted symbol $d_k$ can be modeled as
\begin{equation} \label{Eq:TimeInvarIORelationship}
\begin{aligned}
x_k = H_{k}d_{k}+n_{k}
\end{aligned}
\end{equation}
where $H_k=\sum_{p=0}^{P-1} {h_{p}e^{-j2\pi f_{k}\tau _{p}}}$ is the channel frequency response and $n_{k}$ is the noise.

If we use a large carrier number in the differential OFDM system to insure that the carrier spacing $\Delta f$ is small enough, the channel response would be regarded to change slowly among carriers, in other words, adjacent carriers are nearly the same, e.g., $H_{k-1}\approx H_{k}$. Thus, the estimation of the transmitted symbol $b_k$ is obtained by performing the differentially coherent detection on received symbols $x_k$ and $x_{k-1}$ as
\begin{equation} \label{Eq:DifferentDete}
\begin{aligned}
\hat{b}_k = \frac{x_k}{x_{k-1}}.
\end{aligned}
\end{equation}
After mapping the estimated symbol $\hat{b}_k$ to the nearest point in the constellation, the final symbol $\tilde{b}_k$ will be obtained, which is exactly the transmitted symbol $b_k$.

However, it is worth noting that this approach is carried out under the assumption that the channel variation is negligible. Actually, due to the fact that the acoustic speed is only about 1500 m/s, the motion of the transmitter or receiver, even only at a velocity of 1 m/s, will result a severe Doppler factor, up to the order of $10^{-4}$. In other words, this channel variation cannot be negligible, leading to the conventional differentially coherent detection with a single FFT cannot work well in the real UWA environment with violent Doppler shifts.

\subsection{PS-FFT Demodulation} \label{Subsection:PSFFTDemodulation}
In order to eliminate the ICI mentioned above in differential OFDM systems, we propose a new ICI suppression method, referred to as the PS-FFT, which can be regarded as an extension of the time- and frequency-domain methods, such as the P-FFT and the F-FFT, respectively in \cite{MS_MFFF1_PFFT_Diff_Nomultichan,MS_MFFT3_FFFT_Diff_Multichan}. Compared to the conventional demodulation process described in Section~\ref{Subsection:SingleFFTDemodulation}, we first divide the entire block duration into several short non-overlapping intervals and then apply the Fourier transform to each of the intervals at several predefined frequencies. Fig.~\ref{Fig:PSFFT_Demodulation} illustrates the process of the PS-FFT, and its details are described as follows.

First, we divide the entire block duration $T$ into $A$ short non-overlapping intervals to reduce the channel variation. Moreover, when the divided interval is short enough, the channel variation in each short interval could be negligible. Specifically, this process is equivalent with applying $A$ non-overlapping rectangular windows $\phi _{a}(t)$, defined as (\ref{Eq:TimeWindowFunction}), to the time-domain signal $v(t)$.
\begin{equation} \label{Eq:TimeWindowFunction}
\phi _{a}(t) = {\rm rect}\left ( \frac{At}{T}-\frac{1}{2}-a\right ),\ \ a=0,1,\ldots,{A-1}
\end{equation}
where ${\rm rect}(t)$ is the rectangular window function with a value of $1$ for $t \leq \left | \frac{1}{2}\right |$ and $0$ for other cases. Thus, the windowed signal $v_a(t)$ is given by
\begin{equation} \label{Eq:ReceiveSignalPartial}
v_a(t) = v(t)\phi _{a}(t),\ \ a=0,1,\ldots,{A-1}.
\end{equation}

Next, besides only performing the Fourier transform at the carrier frequency $k\Delta f$ for each carrier, we also perform at other $2L$ frequencies around $k \Delta f$ to compensate Doppler shifts. The $(2L+1)$ frequencies in total are represented as
\begin{equation} \label{Eq:FFTFrequencies}
f_l=k\Delta f+\frac{l}{L+1}f_e,\ \ l=0,\pm 1,\pm 2,\ldots,\pm L
\end{equation}
where $f_e$ is a specific frequency and the offset between adjacent frequencies is $\frac{1}{L+1}f_e$.

After performing the Fourier transform to each of the windowed signals $v_a(t)$ at $(2L+1)$ frequencies $f_l$, the $(2L+1)$ demodulator outputs $z_{k,a,l}$ at each carrier are obtained as
\begin{equation} \label{Eq:XFFTDemodulation}
\begin{aligned}
z_{k,a,l}
&= \int_{0}^{T}{v_a(t)e^{-j2\pi f_{l}t}dt} \\
&= \int_{0}^{T}{v_a(t)e^{-j2\pi \frac{l}{L+1}f_{e}t}e^{-j2\pi k\Delta {f}t}dt}, \\
a&=0,1,\ldots,{A-1}, l=0,\pm 1,\ldots,\pm L.
\end{aligned}
\end{equation}
Since the complex exponential term $e^{-j2\pi \frac{l}{L+1}f_{e}t}$ in (\ref{Eq:XFFTDemodulation}) represents the phase rotation introduced by the frequency $-\frac{l}{L+1}f_e$, we can denote the windowed signal after frequency shifting $v_{a,l}(t)$ as
\begin{equation} \label{Eq:ReceiveSignalPartialFractional}
v_{a,l}(t)=v_{a}(t)e^{-j2\pi \frac{l}{L+1}f_{e}t},\ \ l=0,\pm 1,\ldots,\pm L.
\end{equation}
The demodulation process in (\ref{Eq:XFFTDemodulation}) now can be regarded as performing the Fourier transform at the carrier frequency $k\Delta f$ to each of the $A(2L+1)$ signals $v_{a,l}(t)$.

In order to combine $A(2L+1)$ outputs $z_{k,a,l}$ at the $k$-th carrier, we first gather $(2L+1)$ outputs with the same interval index $a$ at this carrier, and arrange them as
\begin{equation} \label{Eq:InitialVect}
\begin{aligned}
{\bf z}_{k,a}=[z_{k,a,-L},\ldots,z_{k,a,-1},z_{k,a,0},
z_{k,a,1},\ldots,z_{k,a,L}]^{T}.
\end{aligned}
\end{equation}
Then the demodulator output vector ${\bf z}_{k}$ is formed as
\begin{equation} \label{Eq:FinalVect}
\begin{aligned}
{\bf z}_{k}=[{\bf z}_{k,0}^{T},{\bf z}_{k,1}^{T},\ldots,{\bf z}_{k,A-1}^{T}]^{T}.
\end{aligned}
\end{equation}

Finally, a combiner with a weight vector ${\bf w}_{k}$ is applied to ${\bf z}_{k}$, yielding the combined demodulator symbol $x_k$ as
\begin{equation} \label{Eq:CombineOutput}
x_{k}={\bf w}_{k}^{H}{\bf z}_{k}.
\end{equation}

\subsection{Differentially Coherent Detection} \label{Subsection:PSFFTDemodulation}
To arrive the combiner weight ${\bf w}_{k}$, we formulate the optimization problem based on minimizing the MSE as
\begin{equation} \label{Eq:MSEEquation}
\min \, \left | e_k\right |^{2} = \min \, \left | \tilde{b}_k - \hat{b}_k \right |^{2}
\end{equation}
where $\hat{b}_k$ is the estimated symbol after differentially coherent detection, calculated by
\begin{equation} \label{Eq:DiffDetection}
\hat{b}_k = \frac{x_k}{x_{k-1}} = \frac{{\bf w}_{k}^{H}{\bf z}_{k}}{{\bf w}_{k-1}^{H}{\bf z}_{k-1}}.
\end{equation}
Since the carrier spacing is small enough, we can assume that the combiner weights between adjacent carriers are approximately equal, i.e., ${\bf w}_{k-1}^{H}\approx {\bf w}_{k}^{H}$. We thus rewrite the optimization problem in (\ref{Eq:MSEEquation}) as
\begin{equation} \label{Eq:MSEEquation1}
\min \, \left | e_k\right |^{2} = \min \, \left | \tilde{b}_k - \frac{{\bf w}_{k}^{H}{\bf z}_{k}}{{\bf w}_{k}^{H}{\bf z}_{k-1}} \right |^{2}.
\end{equation}

Calculating the partial derivative of the MSE, we obtain the squared error gradient
\begin{equation} \label{Eq:GradientEquation1}
{\bf g}_{k}=-{\partial\{\vert e_{k}\vert ^{2}\}\over \partial {\bf w}_{k}^{\ast}} ={1\over (x_{k-1})^{2}}[{\bf z}_{k}\cdot x_{k-1}-x_{k}\cdot {\bf z}_{k-1}]e_{k}^{\ast}.
\end{equation}
We then employ the SGA with a step size $\mu $ to solve the combiner weights recursively as
\begin{equation} \label{Eq:WeightUpdate}
{\bf w}_{k+1}={\bf w}_{k}+\mu {\bf g}_{k}.
\end{equation}

To carry out the SGA, we divide the solving problem into training mode and decision-directed mode, aiming to start the algorithm and decide transmitted symbols, respectively. Specifically, during the training mode, a total of $N_p$ pilots are inserted in the first few OFDM blocks and the decision symbol $\tilde{b}_k$ is equal to the original symbol $b_k$. After that, the algorithm switches to the decision-directed mode, where $\tilde{b}_k$ is arrived by making decision on $\hat{b}_k$ as $\tilde{b}_k = {\rm{dec}}\left ( \hat{b}_k\right )$.

Moreover, we adopt the gradient scaling method introduced in \cite{MS_MFFT5_MFFT_Diff_Multichan} to enhance the robustness of the weight updating algorithm. Since the denominator in (\ref{Eq:GradientEquation1}) contains the term $(x_{k-1})^{2}$, when the magnitude of the term $x_{k-1}$ is low, the SGA will suffer from noise enhancement. Therefore, we scale the original gradient by term $\left | x_{k-1}\right |$, as in (\ref{Eq:ScaledGradientEquation1}), and use the scaled gradient $\bar{{\bf g}}_{k}$ to update weights.
\begin{equation} \label{Eq:ScaledGradientEquation1}
\bar{{\bf g}}_{k}=\left | x_{k-1}\right |{\bf g}_{k}.
\end{equation}

In addition, we set two error thresholds in the weight updating algorithm to further improve the performance. When the absolute value of the error $e_k$ and the inner product of the gradient ${\bf g}_{k}$ are both smaller than these two thresholds, respectively, the algorithm updates the combiner weights. Otherwise, the weights will not be updated. The weight updating algorithm is summarized in Algorithm~\ref{Algorithm:AdaptiveAlgorithm}.

\begin{algorithm}[!t]
\caption{Weight updating algorithm.}
\begin{algorithmic}[1] \label{Algorithm:AdaptiveAlgorithm}
\STATE $\bf{Initialization}$
\begin{itemize}
  \item Set weight ${\bf w}_{\rm{temp}}$, threshold $e_{\rm{th}}$, $g_{\rm{th}}$, and step size $\mu $.
  \item Set ${\rm {flag}}=+1$.
\end{itemize}
\FOR {all OFDM blocks in one frame}
\STATE Calculate ${\bf z}_{k}$ corresponding to the current block.
\IF {${\rm {flag}}==1$}
\STATE $k=0$.
\ELSE
\STATE $k=K-1$.
\ENDIF
\STATE ${\bf w}_{k}={\bf w}_{k+{\rm{flag}}}={\bf w}_{\rm{temp}}$.
\STATE $x_{k}={\bf w}_{k}^{H}{\bf z}_{k}$.
\STATE $k=k+\rm {flag}$.
\WHILE {$k\leq K-1$ or $k\geq 0$}
\STATE $x_k={\bf w}_{k}^{H}{\bf z}_{k}$.
\STATE $\hat{b}_k={x_k}/{x_{k-{\rm{flag}}}}$.
\IF {pilots available}
\STATE $\tilde{b}_k={\rm{pilot\;symbol}}$.
\ELSE
\STATE $\tilde{b}_k={\rm{dec}}\left ( \hat{b}_k\right )$.
\ENDIF
\STATE $e_k=\tilde{b}_k-\hat{b}_k$.
\IF {$\left | e_k\right |<e_{\rm{th}}$ and ${\bf g}_{k}^{H}{\bf g}_{k}<g_{\rm{th}}$}
\STATE $\mathbf{g}_k=({\bf z}_{k}x_{k-{\rm{flag}}}-{x_{k}\bf z}_{k-{\rm{flag}}})e_{k}^{\ast}/(x_{k-{\rm{flag}}})^2$.
\STATE $\bar{{\bf g}}_{k}=\left | x_{k-{\rm{flag}}}\right |{\bf g}_{k}$.
\STATE ${\bf w}_{k+{\mathrm{flag}}}={\bf w}_{k}+\mu \bar{{\bf g}}_{k}$.
\ELSE
\STATE ${\bf w}_{k+{\mathrm{flag}}}={\bf w}_{k}$.
\ENDIF
\STATE $k=k+\rm {flag}$.
\ENDWHILE
\STATE ${\rm{flag}}=-{\rm{flag}}$.
\STATE ${\bf w}_{\rm{temp}}={\bf w}_{k}$.
\ENDFOR
\end{algorithmic}
\end{algorithm}

\section{Simulation Results and Analysis} \label{Section:SimulationResults}
In this section, we first describe the simulation parameter settings, and then present numerical simulation results to evaluate the performance of the PS-FFT by comparing it with other methods, including the P-FFT and the F-FFT in \cite{MS_MFFT5_MFFT_Diff_Multichan} and the conventional differentially coherent detection with a single FFT (single-FFT), in terms of the MSE.
\subsection{Simulation Parameter Settings} \label{Subsection:SimulationParaSet}
\begin{figure}[t]
\centering \leavevmode \epsfxsize=3.5 in  \epsfbox{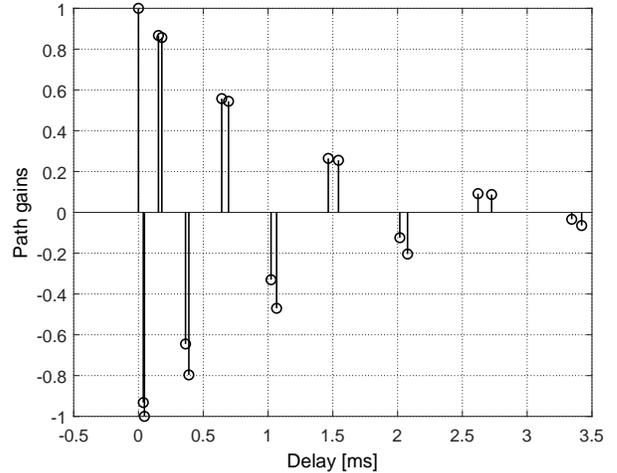}
\centering \caption{Path gains for the considered UWA channel. In this figure, the UWA parameter settings are 15 m water depth, 2 km horizontal distance, 7.5 m transmitter height, 7.5 m receiver height, soft bottom (sound speed 1400 m/s), spreading factor of 1.5 and the center frequency of 32 kHz.} \label{Fig:PathGain}
\end{figure}

In following simulations, we consider a differential OFDM system with typical parameter settings provided in \cite{MS_MFFT2_PFFT_Nodiff_Nomultichan}. Fig.~\ref{Fig:PathGain} illustrates path gains of the UWA channel, which are calculated by the statistical channel model in \cite{MS_UWA_ChannelModel}. The detailed OFDM parameter settings are summarized in Table~\ref{Table:OFDMParameters}, in which the number of carriers $K$ and the number of blocks per frame $N$ are varying and satisfied with $KN=2^{13}$.

For PS-FFT, P-FFT, and F-FFT methods, we adopt the SGA equipped with the gradient scaling and the thresholding method proposed in \cite{MS_MFFT5_MFFT_Diff_Multichan} to solve the combiner weights. The number of divided intervals for P-FFT and the number of Fourier transform frequencies for F-FFT are $A=3$ and $(2L+1)=3$, respectively. The PS-FFT uses the same parameters $A=3$ and $(2L+1)=3$ as the P-FFT and F-FFT. Noting, the values of the Doppler factor in following simulations are residuals after initial resampling, which are equal to the ratio of the frequency shift to the carrier frequency $f_k$. Since the maximum Doppler factor in following simulations is $\alpha _{\rm{max}}=3\times 10^{-4}$, corresponding to a Doppler shift at the center carrier $f_{d,{\rm{max}}}=\alpha _{\rm{max}} f_c=9.6$ Hz, it is reasonable to set the parameter of the fiducial offset frequency $f_e$ for PS-FFT as $f_e=2f_{d,{\rm{max}}}=19.2$ Hz. In addition, we insert a total of $N_p=250$ pilots in the first few blocks to start the SGA.

\begin{table}[t]
\centering
\caption{\label{Table:OFDMParameters}Summary of the UWA-OFDM parameter settings.}
\begin{supertabular}{l >{}p{3.1cm}}
\shline
Parameters & Values  \\
\hline
Center frequency $f_c$      & 32 kHz\\
\hline
Signal bandwidth $B$        & 12 kHz\\
\hline
Sampling rate $f_s$        & 192 kHz\\
\hline
Sampling interval $T_s$       & 5.208 us\\
\hline
Number of carriers per block $K$   & $2^6 - 2^{11}$\\
\hline
Number of blocks per frame $N$     & $2^2 - 2^7$\\
\hline
Carrier spacing $\Delta f$  & $5.9 - 187.5$ Hz\\
\hline
Block duration $T$          & $5.3 - 170.7$ ms\\
\hline
Guard interval $T_g$        & 16 ms\\
\hline
Modulation type             & QPSK\\
\shline
\end{supertabular}
\end{table}

\subsection{MSE Versus Doppler Factor} \label{Subsection:MSEPerformanceVersusDoppler}
\begin{figure}[t]
\centering \leavevmode \epsfxsize=3.5 in  \epsfbox{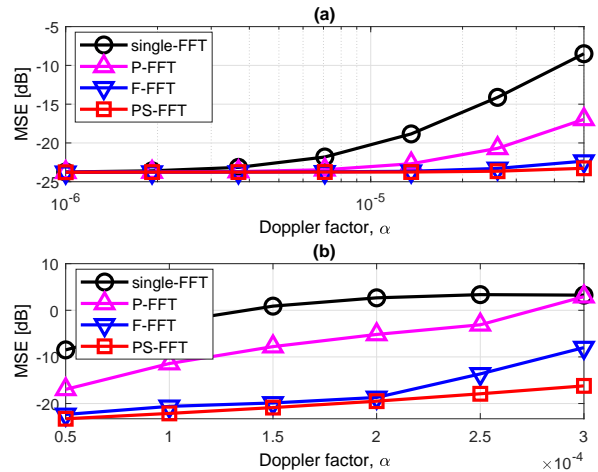}
\centering \caption{Comparisons of the MSE versus the Doppler factor $\alpha$ among different methods, where $K=1024$ and $\rm{SNR}=30$ dB. (a) $\alpha$ ranges from $10^{-6}$ to $5\times 10^{-5}$. (b) $\alpha$ ranges from $5\times 10^{-5}$ to $3\times 10^{-4}$.} \label{Fig:MSEversusDoppler}
\end{figure}

In Fig.~\ref{Fig:MSEversusDoppler}, we compare the MSE performance of different methods in terms of the Doppler factor $\alpha$ with the number of carriers $K=1024$ and the input signal-to-noise ratio (SNR) at the receiver of 30 dB. Moreover, Fig.~\ref{Fig:MSEversusDoppler}(a) compares in the low Doppler factor interval where $\alpha =10^{-6}- 5\times 10^{-5}$, while Fig.~\ref{Fig:MSEversusDoppler}(b) in the medium and high interval where $\alpha =5\times 10^{-5} - 3\times 10^{-4}$. From Fig.~\ref{Fig:MSEversusDoppler}(b), it can be found that the conventional method without ICI suppression cannot work when $\alpha $ achieves a large value, i.e., $10^{-4}$, while other methods keep good performance at this point. Among the three methods with multiple demodulator outputs, the F-FFT outperforms P-FFT, revealing that mitigating the ICI from the frequency domain is more efficient than the time domain. Furthermore, the PS-FFT, mitigating the ICI from both the time and frequency domains, dramatically outperforms F-FFT at a medium or high Doppler factor. Numerically, the MSE of the PS-FFT is reduced by $84.89\%$ compared to the F-FFT when $\alpha =3\times 10^{-4}$ and $\rm{SNR}=30$ dB.

\subsection{MSE Versus the Number of Carriers} \label{Subsection:MSEPerformanceVersusCarrierNum}
\begin{figure}[t]
\centering \leavevmode \epsfxsize=3.5 in  \epsfbox{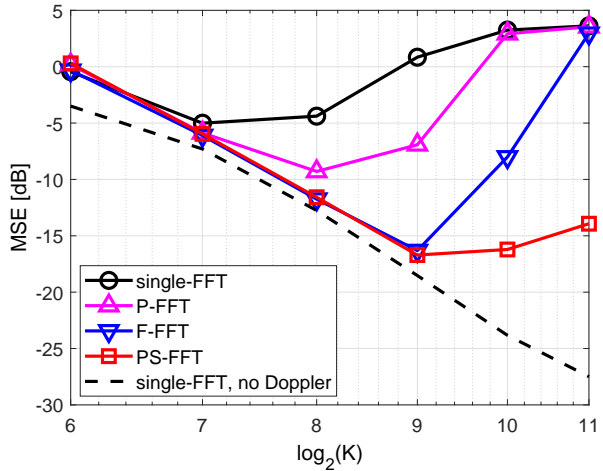}
\centering \caption{Comparisons of the MSE versus the number of carriers $K$ among different methods, where $\rm{SNR}=30$ dB and $\alpha=3\times 10^{-4}$ for the top four solid curves and $\alpha=0$ for the bottom dotted curve.} \label{Fig:MSEversusCarrierNum}
\end{figure}

Fig.~\ref{Fig:MSEversusCarrierNum} demonstrates the MSE performance in terms of the number of carriers $K$ with $\rm{SNR}=30$ dB and Doppler factor $\alpha=3\times 10^{-4}$. Besides, the performance of the conventional method with no Doppler is also provided, which is monotonically improved as $K$ getting larger and acts as a performance upper bound. It can be found that the performance of each method is improved as $K$ increasing from 64 to 128. This can be explained by increasing the number of carriers, the carrier spacing becomes narrower and thus enhances the coherence between adjacent carriers. However, shortening the carrier spacing $\Delta f$ makes the system quite sensitive to the frequency offset, resulting that the performance of the conventional method begins to drop at $K=128$. For other methods possessed with ICI mitigation, this decline trend can be postponed. Furthermore, it is worth noting that as $K$ achieving 2048, neither P-FFT nor F-FFT can work, but the PS-FFT still maintains MSE performance at about $-14$ dB.

\subsection{MSE Versus SNR} \label{Subsection:MSEPerformanceVersusSNR}
\begin{figure}[t]
\centering \leavevmode \epsfxsize=3.5 in  \epsfbox{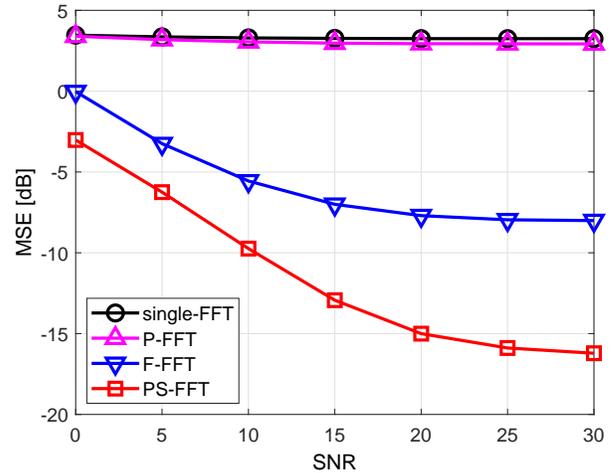}
\centering \caption{Comparisons of the MSE versus SNRs among different methods, where $K=1024$ and $\alpha= 3\times 10^{-4}$.} \label{Fig:MSEversusSNR}
\end{figure}

Fig.~\ref{Fig:MSEversusSNR} compares the MSE performance of different methods in terms of the input SNR at the receiver with the number of carriers $K=1024$ and Doppler factor $\alpha=3\times 10^{-4}$. For the conventional method, the performance is not improved as the increase of the SNR, which means that the ICI cannot be compensated by increasing the SNR. Since the MSE performance of the P-FFT is above 0 dB and cannot work as the conventional method, we only concern the F-FFT and the PS-FFT. Both methods achieve better performance as the increase of the SNR by eliminating ICI, and the PS-FFT demodulation reduces the MSE by $61.83\%-84.89\%$ compared to the F-FFT when the SNR ranges from 10 dB to 30 dB.

\section{Conclusions} \label{Section:Conclusions}
In this paper, we have proposed a ICI suppression method referred to as the PS-FFT, eliminating the ICI from both the time and frequency domains, to improve the performance of differentially coherent detection when the Doppler factor and the number of carriers are both large. Specifically, it first divides the block duration into several short non-overlapping intervals for channel variation reduction, and then performs the Fourier transform to the received signal in each of these interval at several predefined frequencies to compensate Doppler shifts. Finally, it utilizes the SGA to solve the weights for combining the multiple demodulator outputs. Simulation results have verified that the performance of the PS-FFT is significantly superior to the existing methods, the P-FFT and the F-FFT, for both medium and high Doppler factors and large carrier numbers.

\bibliographystyle{IEEEtran}
\bibliography{IEEEabrv,LatexWritingModel_Reference}

\end{document}